\documentclass[aps,prl,reprint,superscriptaddress,nofootinbib,floatfix]{revtex4-2}

\usepackage{amsmath,amssymb,bm}
\usepackage{graphicx}
\usepackage{hyperref}
\usepackage{microtype}

\newcommand{\ii}{\mathrm i}
\newcommand{\cw}{c_{w}} 
\newcommand{\rr}{r_0}
\newcommand{\kap}{\kappa}
\newcommand{\Gv}{\bm G}

\usepackage[normalem]{ulem}  
\usepackage{xcolor}

\begin{document}

\title{Exact Rotating Wormholes in Five Dimensions and Their Exceptional-Point Instability}

\author{Jose Luis Bl\'azquez-Salcedo}
\email{jlblaz01@ucm.es}
\affiliation{Departamento de F\'isica Te\'orica and IPARCOS, Facultad de Ciencias F\'isicas, Universidad Complutense de Madrid, 28040 Madrid, Spain}

\author{Luis Manuel Gonz\'alez-Romero}
\email{mgromero@ucm.es}
\affiliation{Departamento de F\'isica Te\'orica and IPARCOS, Facultad de Ciencias F\'isicas, Universidad Complutense de Madrid, 28040 Madrid, Spain}

\author{Fech Scen Khoo}
\email{fkhoo@ucm.es}
\affiliation{Departamento de F\'isica Te\'orica and IPARCOS, Facultad de Ciencias F\'isicas, Universidad Complutense de Madrid, 28040 Madrid, Spain}

\author{Burkhard Kleihaus}
\email{b.kleihaus@uni-oldenburg.de}
\affiliation{Institut f\"ur Physik, Universit\"at Oldenburg, Postfach 2503, D-26111 Oldenburg, Germany}

\author{Jutta Kunz}
\email{jutta.kunz@uni-oldenburg.de}
\affiliation{Institut f\"ur Physik, Universit\"at Oldenburg, Postfach 2503, D-26111 Oldenburg, Germany}

\author{Pablo Navarro Moreno}
\email{panava03@ucm.es}
\affiliation{Departamento de F\'isica Te\'orica and IPARCOS, Facultad de Ciencias F\'isicas, Universidad Complutense de Madrid, 28040 Madrid, Spain}

\date{\today}

\begin{abstract}
We give an independent closed-form construction of rotating five-dimensional Ellis--Bronnikov wormholes with equal angular momenta and analyze the radial instability of the reflection-symmetric family.  
The nontrivial coupled metric equations reduce exactly to an open $A_2$ Toda system. 
The general two-ended solution is expressed in terms of elementary Toda tau functions; the symmetric subclass reduces to a particularly simple one-parameter family interpolating between the static wormhole and the extremal equal-spin Myers--Perry geometry.  
On this exact background we revisit the cohomogeneity-one perturbation problem.  
Besides the ordinary unstable mode of the static wormhole, rotation renders the descendant of its non-normalizable zero mode normalizable.  
The two purely imaginary unstable frequencies coalesce at $|J|/A=0.100628$, where their eigenvectors coalesce as well, identifying a second-order exceptional point.  
Beyond it the pair continues as $\omega_\pm=\pm\omega_R+\ii\omega_I$ with $\omega_I>0$: 
the apparent loss of the radial instability found in the original real-eigenvalue treatment
is therefore not stabilization, but a transition to an oscillatory instability.  
We follow this branch deep into the rapidly rotating regime, where its growth rate becomes
strongly suppressed.  
\end{abstract}

\maketitle

\paragraph{Introduction.}
The Ellis--Bronnikov wormhole \cite{Ellis:1973yv,Bronnikov:1973fh} is the canonical asymptotically flat traversable wormhole supported by a phantom scalar field.
The static Ellis–Bronnikov wormhole is, however, known to be linearly unstable against spherically symmetric perturbations: in four dimensions it possesses a single exponentially growing radial mode, as established by several complementary formulations \cite{Gonzalez:2008wd,Bronnikov:2011if,Cremona:2018wkj,Cremona:2019wiy}, while nonlinear evolutions lead either to collapse or to an expanding wormhole configuration \cite{Shinkai:2002gv,Gonzalez:2008xk}. 
The instability persists for the higher-dimensional generalizations, which likewise possess a radial unstable mode \cite{Torii:2013xba}. 

Rotating generalizations of the Ellis--Bronnikov wormhole were first studied perturbatively in the slow-rotation regime \cite{Kashargin:2007mm,Kashargin:2008pk}, while fully nonlinear rotating solutions were subsequently constructed numerically \cite{Kleihaus:2014dla,Chew:2016epf}. 
The latter form continuous families connecting the static Ellis--Bronnikov wormhole to the extremal Kerr limit. 
This raised the question whether rotation might also weaken or remove the radial instability, which was addressed perturbatively for slowly rotating wormholes \cite{Azad:2023iju,Azad:2024axu}.
More recently, quasinormal modes of rapidly rotating wormholes were calculated in nonradial sectors \cite{Khoo:2024yeh}, while fully nonlinear time evolutions have confirmed that rotation substantially slows the instability \cite{Shirokov:2026otu}.

In five dimensions, imposing equal angular momenta enhances the rotational
symmetry sufficiently to preserve cohomogeneity one.  
The resulting rotating wormholes were constructed numerically in Ref.~\cite{Dzhunushaliev:2013jja}, where their approach to the extremal equal-spin Myers--Perry black hole \cite{Myers:1986un} and their radial stability were also studied.  
More recent numerical work explored strongly asymmetric configurations \cite{Uemichi:2026dzb}.

Here we report two results.  
First, the complete stationary equations admit an exact Toda reduction and closed-form solution.  
Second, using this exact background we resolve the fate of the two radial unstable modes of the reflection-symmetric family.  
The merger previously interpreted as the disappearance of the instability is an exceptional point (EP): 
when the spectral problem is continued into the complex plane, the two modes persist as an oscillatory unstable pair.  

Exceptional points have recently been identified in black-hole quasinormal-mode spectra, including massive scalar perturbations of Kerr and scalar or gravito-electromagnetic perturbations of hairy and Kerr–Newman black holes. 
These examples concern damped quasinormal modes \cite{Cavalcante:2024swt,Oshita:2025ibu,Cheng:2026gxu,Arcos:2026eis}. 
The associated resonant excitation and non-Hermitian mode interaction near EPs have also been investigated in the context of black-hole ringdown \cite{Motohashi:2024fwt,Yang:2025dbn,PanossoMacedo:2025xnf,Kubota:2025hjk}.
To the best of our knowledge, the present case provides the first explicit identification, including eigenvector coalescence, of an exceptional point formed by two dynamically unstable modes, and the first such identification in the perturbation spectrum of a fully self-gravitating horizonless compact object.

Closely analogous spectral behavior was already found for supercritical charged Ellis–Bronnikov wormholes, where two purely imaginary unstable radial modes merge and subsequently acquire opposite real parts while retaining the same growth rate. 
Although this merger was not interpreted in terms of exceptional-point topology in Refs.~\cite{Blazquez-Salcedo:2025dit,Blazquez-Salcedo:2026cps}, it can now be recognized as the same characteristic spectral mechanism.

\paragraph{Theory.}

The wormholes are described by Einstein gravity coupled to a phantom scalar field $\phi$ in five dimensions,
\begin{equation}
    S = \int \left[\frac{1}{16\pi G} \mathcal{R} + \mathcal{L}_{\phi} \right] 
    \sqrt{-g} \, d^5x \, ,
\end{equation}
where $\mathcal{L}_{\phi}=\frac{1}{2}\partial_{\mu}\phi\partial^{\mu}\phi$.
The field equations are, 
for the metric,
\begin{equation}
   G_{\mu\nu} = \mathcal{R}_{\mu\nu} - \frac{1}{2} g_{\mu\nu}\mathcal{R} = 8\pi GT_{\mu\nu} \, ,
   \label{meEq}
\end{equation}
where
$T_{\mu\nu}
=g_{\mu\nu}\mathcal{L}_{\phi} - 2 \, 
\partial\mathcal{L}_{\phi} / \partial g^{\mu\nu}$, and for the phantom field,
\begin{equation}
    \nabla_{\mu}\nabla^{\mu} \phi = 0\,.
    \label{scEq}
\end{equation}

\paragraph{Exact rotating wormholes.}

The rotating background ansatz is given by
\begin{eqnarray}
    ds^2 =
    & -e^{2a({r})} dt^2 +
    p({r}) e^{-q({r})} \left(d{r}^2+ \frac{1}{4} h({r}) (\sigma_1^2 + \sigma_2^2) \right) 
    \nonumber \\ &
    + \frac{1}{4} p({r}) e^{q({r})-2a({r})} h({r}) \left(\sigma_3 -2 w({r}) dt \right)^2 \,,
\end{eqnarray}
where $\sigma_i$'s are the left-invariant 1-forms on $S^3$,
while for the phantom field it is simply
\begin{equation}
    \phi = \phi({r}) \,.
\end{equation}
We use the radial gauge of Ref.~\cite{Dzhunushaliev:2013jja},
\begin{equation}
 h={r}^2+\rr^2 \,,\qquad
 p=\frac{{r}^2+\rr^2/2}{{r}^2+\rr^2} \,,
\end{equation}
for which the nontrivial stationary functions are $a({r})$, $q({r})$ and $w({r})$.  
In order to obtain an analytical solution it is useful to define the transformation
\begin{equation}
 u=\arctan\!\left(\frac{{r}}{\sqrt{{r}^2+\rr^2}}\right) \,,
 \qquad -\frac{\pi}{4}<u<\frac{\pi}{4} \,.
 \label{eq:u}
\end{equation}
Then it is possible to see that the field equations can be simplified if we define the function $b$ and the parameter $\kappa$:
\begin{equation}
 b=q-\ln(\cos2u) \,,\qquad \kap=\frac{\cw}{\rr^3} = -\frac{16 G}{\pi r_0^3}J \,,
\end{equation}
where $J=J_1=J_2$ is the angular momentum.
Then the two nonlinear Einstein equations 
become
\begin{equation}
 a_{uu}=4\kap^2 e^{4a-b} \,,\qquad
 b_{uu}=4e^{2(b-a)} \,.
\end{equation}
Even more, defining the functions
\begin{equation}
 U=4a-b+\ln(8\kap^2)\,,\qquad V=2(b-a)+\ln4 \,,
\end{equation}
one obtains the open $A_2$ Toda system
\begin{equation}
 U_{uu}=2e^U-e^V \,,\qquad V_{uu}=-e^U+2e^V \,.
 \label{eq:toda}
\end{equation}

A real sector relevant to regular two-ended wormholes is generated by the Toda
exponents
$\{2\alpha,-\alpha+\ii\beta,-\alpha-\ii\beta\}$.  
Introducing
\begin{align}
 \tau_1={}&e^{2\alpha u}+2B e^{-\alpha u}\cos(\beta u+\delta)\,,\\
 \tau_2={}&4B^2\beta^2e^{-2\alpha u}
 -2Be^{\alpha u}\Big[(9\alpha^2-\beta^2)\cos(\beta u+\delta)
 \nonumber\\[-1mm]
 &\hspace{27mm}+6\alpha\beta\sin(\beta u+\delta)\Big]\,,
 \label{eq:taus}
\end{align}
and
$\mathcal C=4B^2\beta^2(9\alpha^2+\beta^2)^2$, the metric functions are
\begin{equation}
 e^{2a}=\frac{\mathcal C^{1/3}}
 {(16\kap^2)^{2/3}\tau_1} \,,\qquad
 e^q=\frac{\mathcal C^{2/3}}{2(16\kap^2)^{1/3}}
 \frac{\cos(2u)}{\tau_2} \,,
 \label{eq:generalmetric}
\end{equation}
while
\begin{equation}
 w
 =-\frac{1}{2\kap\rr}
 \left(\frac{\tau_1'}{\tau_1}
 -\frac{\tau_1'(\pi/4)}{\tau_1(\pi/4)}\right) \,,\qquad
 \phi=\frac{2Q}{\rr^2}\left(u-\frac{\pi}{4}\right) \,.
 \label{eq:omegaphi}
\end{equation}
These functions depend on the parameters $\alpha$, $\beta$, $\delta$, $\kappa$, $B$, $r_0$ and $Q$, but for a physical solution they are not all free.
Asymptotic flatness requires $\tau_2(\pm\pi/4)=0$, which fixes $B$ and $\delta$ algebraically.
The normalization at
positive infinity gives
\begin{equation}
\kap^2=\frac{\sqrt{\mathcal C}} {16\,\tau_1(\pi/4)^{3/2}} \,,
\end{equation}
and the remaining Einstein constraint fixes the scalar charge
\begin{equation}
 4\pi GQ^2=\frac{\rr^4}{16}(16+3\alpha^2-\beta^2) \,.
 \label{eq:Qgeneral}
\end{equation}
Thus physical solutions depend on three free parameters, $(\rr,\alpha,\beta)$
which are related to the throat area, the angular momentum and the mass of the wormhole.

The general solution is generically asymmetric under the reflection $r\to -r$. Let us now focus on the symmetric subclass. Such solutions satisfy
\begin{equation}
\alpha=0\,,\quad \beta=\lambda\,, \quad\delta = 0 \,,\quad 2\leq\lambda<4 \,.
\end{equation}
In fact it is convenient to define $s=\frac{\pi\lambda}{4}$.%
The solution simplifies to
\begin{align}
 e^{2a}&=\frac{\sin^2(s)}{1-\cos(s)\cos(\lambda u)}\,,\\
 e^q&=\frac{\lambda\sin(s)}{2}\,
 \frac{\cos(2u)}{\cos(\lambda u)-\cos(s)}\,,
\end{align}
with
\begin{align}
 \kap^2=-\frac{\lambda^3\cos(s)}{16\sin^3(s)},\qquad
 4\pi GQ^2=\rr^4\left(1-\frac{\lambda^2}{16}\right)\,.
\end{align}
The static five-dimensional Ellis--Bronnikov solution is at 
$\lambda=2$,
while $\lambda\rightarrow4$ sends $Q\rightarrow0$ and yields the extremal equal-spin Myers--Perry limit. 

The hypersurface area defined by
\begin{equation}
    A_T({r}) = 2\pi^2
    \left( 
    ({r}^2 + r_0^2)^3 p^3 e^{-(2a+q)}
    \right)^{1/2} \, ,
\end{equation}
gives the area of the throat $A=A_T(0)$ at the location of the throat, ${r}=0$.
Then it is possible to see that when $\lambda\rightarrow4$ then $|J|/A\rightarrow1/2$, reproducing the numerical limiting behavior of Ref.~\cite{Dzhunushaliev:2013jja} analytically,
where the angular momentum $J$ is identified from the expansion of $w$ at
positive spatial infinity.

\paragraph{Radial perturbations.}
We perturb the metric functions, the scalar field and the angular velocity as in Ref.~\cite{Dzhunushaliev:2013jja}, introducing a harmonic time dependence $e^{-\ii\omega t}$. 
We fix the gauge imposing $H_1=H_3-H_2$. 
With this gauge choice, the phantom-field and angular-velocity perturbations decouple from the gravitational perturbation sector considered here.
It is convenient to introduce the following redefinitions
\begin{equation}
 H_0=G_2+G_0 \,,\quad H_2=-G_2 \,,\quad H_3=2G_0+G_1 \,,
\end{equation}
because with them the perturbation equations can be written in a very simple way:
\begin{equation}
 \frac1{\xi}(\xi\Gv')'=\bm V({r})\Gv-\omega^2F({r})\bm C\Gv\,,
 \,\,
 \bm C=\begin{pmatrix}1&1&0\\0&1&0\\0&0&1\end{pmatrix}\,,
 \label{eq:pert}
\end{equation}
where $\xi=ph^{3/2}$ and $F=e^{-2a-q}p$.  All entries of $\bm V$ are elementary functions of the exact background.

We are interested in perturbations behaving like outgoing waves at both spatial infinities ($r\to\pm\infty$). This implies that the perturbation functions behave like:
\begin{align}
 G_0&\sim e^{i\omega \lvert r\rvert}\lvert r\rvert^{3/2}\,, &
 G_1&\sim e^{i\omega \lvert r\rvert}\lvert r\rvert^{1/2}\,,\nonumber\\[-1mm]
 G_2&\sim e^{i\omega \lvert r\rvert}\lvert r\rvert^{1/2}\,.
 \label{eq:asympt}
\end{align}
With this information we proceed to solve equation (\ref{eq:pert}) numerically in the following way: First
we factor this asymptotic behavior analytically and compactify the radial coordinate according to 
\begin{equation}
 r=r_0\tan\left(\frac{\pi}{2}y(x)\right) \,,
\end{equation}
where $y(x)=x+\frac{\gamma}{\pi}\sin\left(\pi x\right)$, $x\in[-1,1]$ and $\gamma$ is a parameter used to monitor the precision of the numerical method.
We typically choose values $0\le\gamma\le0.8$. 
Then we use a Chebyshev collocation method similar to the one used in other settings \cite{Khoo:2024yeh,Blazquez-Salcedo:2023hwg,Blazquez-Salcedo:2024oek, Khoo:2024agm, Blazquez-Salcedo:2024dur, Berti:2025hly} to calculate the spectrum of quasinormal modes, and in particular, the unstable modes.

\begin{figure}
 \includegraphics[angle=-90,width=\columnwidth]{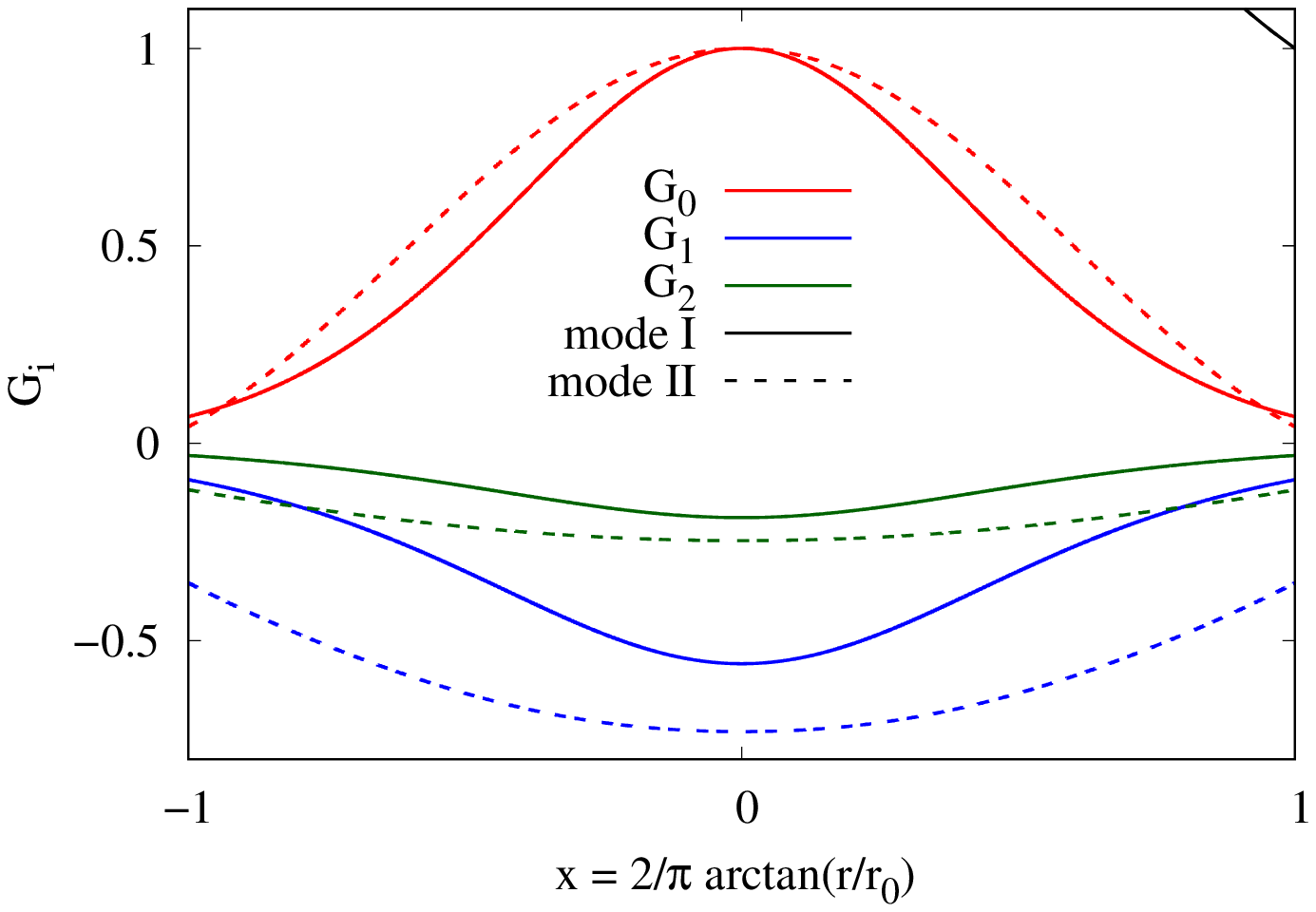}
 \caption{
 Two cohomogeneity-one unstable modes of the symmetric rotating  wormhole at $|J|/A=0.05475$ ($\lambda=2.059523$, $\rr=0.5$).
 The perturbations are normalized by $G_0(0)=1$.  
 The eigenvalues are $\omega^2=-19.1338$ (mode I or deep mode) and $-0.51154$ (mode II or shallow mode). 
 }
  \label{fig:modes}
\end{figure}

As an example, we show in figure~\ref{fig:modes} two unstable eigenfunctions for a background solution with
$|J|/A=0.05475$ and $\rr=0.5$,
already used in Ref.~\cite{Dzhunushaliev:2013jja}.  
We recover mode I with
$\omega^2=-19.1338$ (deep mode) and mode II with $\omega^2=-0.51154$ (shallow mode).
The shallow mode connects with 
the non-normalizable static zero mode
when the angular momentum of the wormhole is decreased.
The figure also illustrates that the two eigenfunctions are distinct well below the EP.

\begin{figure}
 \includegraphics[angle=-90,width=\columnwidth]{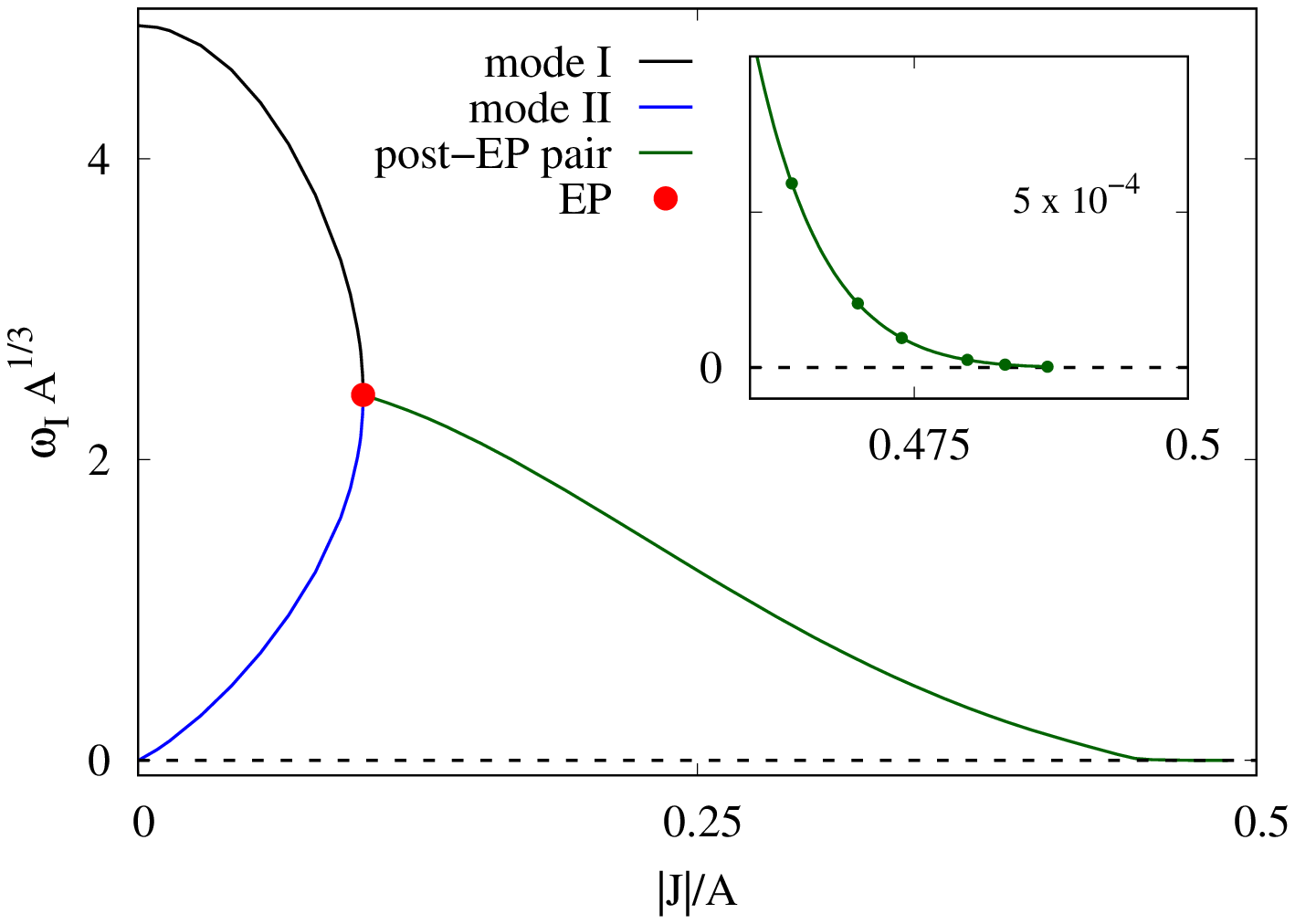}
 \caption{Dimensionless growth rates $\mathrm{Im}(\omega)A^{1/3}$ versus $|J|/A$.  
 The ordinary static instability (mode I) and the zero-mode descendant (mode II) coalesce at the EP.  
 Beyond the EP they form one complex-frequency pair with common positive imaginary part.}
 \label{fig:im}
\end{figure}

\paragraph{Exceptional point and beyond:
Oscillatory instability.}
As shown in figure~\ref{fig:im}, 
the two imaginary frequencies approach each other as the wormhole rotates faster and meet at the EP, where
\begin{eqnarray}
\lambda_{\rm EP}=2.190024 \,,
 \\
 \left(\frac{|J|}{A}\right)_{\rm EP}=0.100628 \,,
 \label{eq:EP}
\end{eqnarray}
and the corresponding frequency is
\begin{eqnarray}
\omega_{\rm EP}  A^{1/3}  =2.429974 \,\ii \,,
 \\
 \omega_{\rm EP}^2  A^{2/3}=-5.904772 \, .
\end{eqnarray} 
The coalescence occurs not only for the eigenvalues but also for the eigenvectors.
Numerically it is possible to see that as $\lambda$ approaches the EP value, the normalized eigenvectors coalesce, becoming more and more similar.
The two eigenvalues exhibit the characteristic square-root branching of an EP2; together with the eigenvector coalescence, this identifies the merger as a second-order EP
\cite{Heiss:2012}.

Immediately beyond the EP the eigenvalues depart from the  real $\omega^2$ axis.  
For example, at 
$\lambda=2.192095$,
\begin{equation}
 \omega_\pm A^{1/3}=\pm0.863525+2.427364\,\ii \,.
\end{equation}
Thus both members of the eigenvalue pair remain unstable.  
The earlier real-eigenvalue calculation \cite{Dzhunushaliev:2013jja} correctly located the merger, but could not follow the modes once $\omega^2$ left the real axis.

\begin{figure}
 \includegraphics[angle=-90,width=\columnwidth]{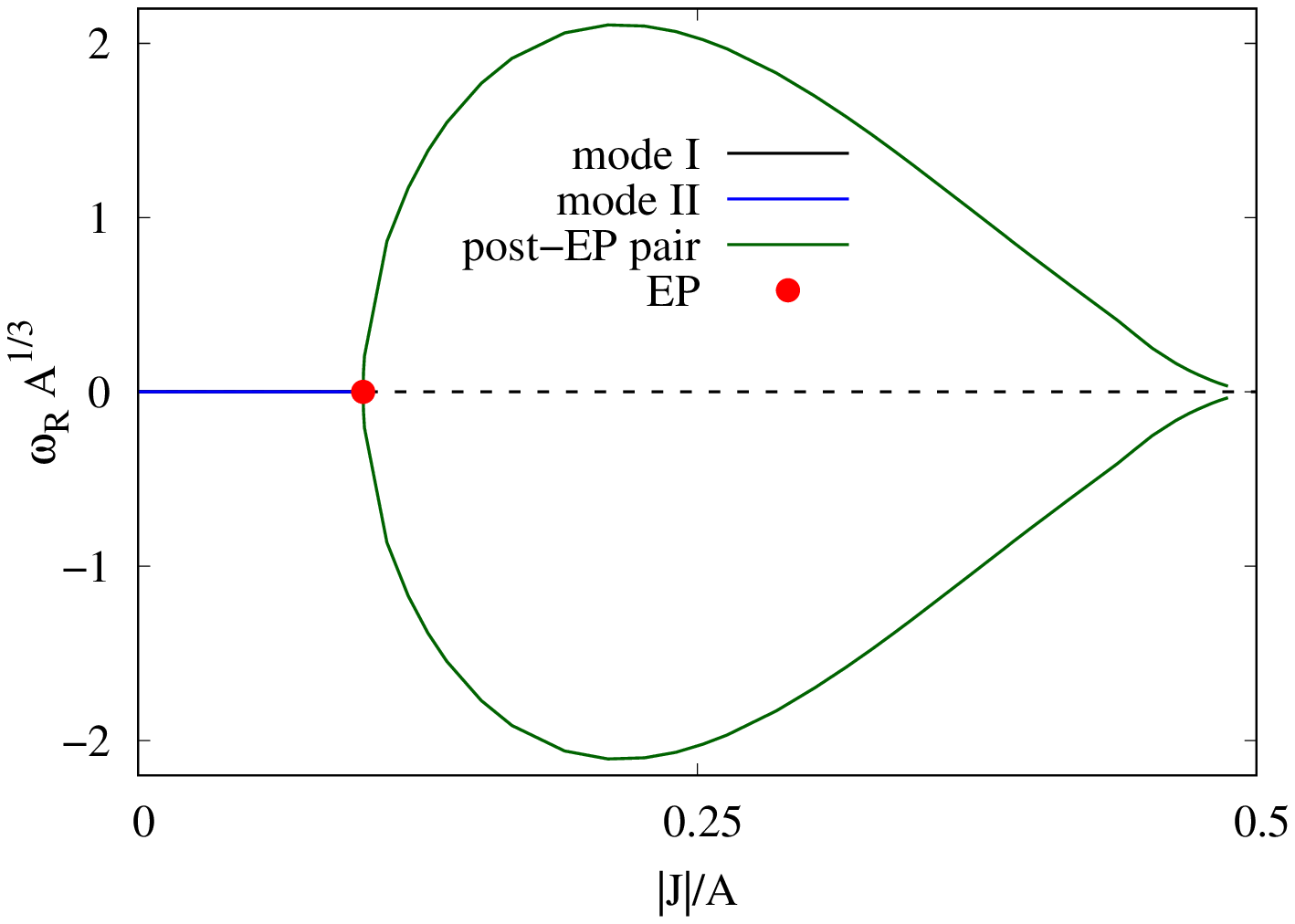}
 \caption{Real parts of the post-EP frequencies.  
 The two branches emerge from the EP with opposite real parts, $\omega_\pm=\pm\omega_R+\ii\omega_I$, demonstrating the transition from a purely exponential to an oscillatory instability.}
 \label{fig:re}
\end{figure}

The evolution of the complex frequency is shown in figures~\ref{fig:im} and \ref{fig:re},
demonstrating an oscillatory instability beyond the EP.
The growth rate of the instability declines rapidly post-EP as the wormhole rotates increasingly fast.
Furthermore, 
the suppression of the instability is very strong, with the dimensionless growth rate $\omega_I A^{1/3}$
reaching values of order $10^{-6}$.
On the other hand, with respect to rotation, the value of the post-EP real frequency rises to a maximum before decreasing closely to zero as well,
where $|J|/A\simeq0.48717$ 
($\lambda=3.948341$). 
Both the increasingly small growth rate and real frequency towards the extremal Myers--Perry limit at $|J|/A=0.5$ suggest a zero-frequency endpoint.
The numerical data are consistent with a behavior
$\omega_I \sim (|J|/A-0.5)^6$ towards the extremal limit.

\paragraph{Discussion.}
The exact Toda solution describing a fully rotating five-dimensional wormhole of equal angular momenta is an intriguing analytical solution. 
The earlier stability analysis of Ref.~\cite{Dzhunushaliev:2013jja}, based on a numerically constructed background, suggested a restoration of radial stability as rapid rotation sets in.
In this work we show that the two unstable modes coalesce at the EP at 
$|J|/A\simeq0.100628$ and proceed to exhibit an oscillatory instability.
This realizes, in an unstable sector, the same non-Hermitian mode-coalescence mechanism that underlies the resonant behavior discussed for black-hole quasinormal modes near exceptional points \cite{Motohashi:2024fwt,Yang:2025dbn,PanossoMacedo:2025xnf,Kubota:2025hjk}.

The endpoint is also structurally interesting.  
The scalar charge vanishes and the background approaches extremal Myers--Perry, while the unstable frequency is driven towards zero.  
It will require a dedicated near-extremal analysis to examine whether the limiting mode is a threshold mode at the wormhole-to-black-hole transition, or is a standard Myers--Perry quasinormal mode.
Nonetheless, our central conclusion remains intact that the symmetric rotating five-dimensional wormholes are linearly unstable in the cohomogeneity-one gravitational sector.

\paragraph{Note added.}
After our exact construction and stability analysis had been obtained, the preprint by Nozawa \cite{Nozawa:2026} appeared, giving an independent exact construction of the rotating five-dimensional wormhole family from an $A_1\oplus A_2$ Toda reduction and analyzing its global geometry.  
The present work is complementary in its emphasis on the radial spectrum, the exceptional point, and the oscillatory instability beyond the merger.

\paragraph{Acknowledgments.}
FSK gratefully acknowledges support from ``Atracci\'on de Talento Investigador Cesar Nombela'' of the Comunidad de Madrid under the grant number 2024-T1/COM-31385.
PNM gratefully acknowledges support from Universidad Complutense de Madrid through ``Contratos predoctorales de personal investigador en formación CT25/24''.

\paragraph{Use of generative AI.}

Generative AI was used as an auxiliary tool in obtaining and organizing parts of the derivation of the exact solutions and for language editing. 
The exact solutions were independently verified by the authors, while the perturbation analysis, derivation of the unstable modes, numerical calculations, and physical interpretation were carried out and checked by the authors.

\end{document}